\documentclass{WileyMSP-template}
\usepackage{booktabs,chemformula}
\usepackage{color,soul}
\usepackage{upgreek}

\begin{document}

\pagestyle{fancy}
\rhead{\includegraphics[width=3cm]{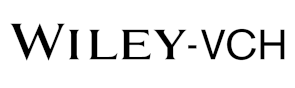}}
\setlength{\parindent}{1em}
\setlength{\parskip}{0.75em}

\title{\noindent Robust electrothermal switching of optical phase change materials through computer-aided adaptive pulse optimization}

\maketitle

\author{Parth Garud$^{1,2}$*},
\author{Kiumars Aryana$^1$*},
\author{Cosmin Constantin Popescu$^3$},
\author{Steven Vitale$^4$},
\author{Rashi Sharma$^5$},
\author{Kathleen Richardson$^5$},
\author{Tian Gu$^{3,6}$},
\author{Juejun Hu$^{3,6}$},
\author{Hyun Jung Kim$^1$}%

\begin{affiliations}
\noindent$^1${NASA Langley Research Center, Hampton, VA 23681-2199, USA}\\
\noindent$^2${Daniel Guggenheim School of Aerospace Engineering, Georgia Institute of Technology, Atlanta, GA 30332, USA}\\
\noindent$^3${Department of Materials Science and Engineering, Massachusetts Institute of Technology, Cambridge, 02139, MA, USA}\\
\noindent$^4${Lincoln Laboratory, Massachusetts Institute of Technology, Lexington, MA 02421, USA}\\
\noindent$^5${The College of Optics and Photonics, Department of Materials Science and Engineering, University of Central Florida, Orlando, FL 32816, USA}\\
\noindent$^6${Materials Research Laboratory, Massachusetts Institute of Technology, Cambridge, 02139, MA, USA}\\

\noindent
*These authors contributed equally.

\noindent
Email: kiumars.aryana@nasa.gov, hyunjung.kim@nasa.gov

\end{affiliations}


\begin{abstract}
\noindent Electrically tunable optical devices present diverse functionalities for manipulating electromagnetic waves by leveraging elements capable of reversibly switching between different optical states. This adaptability in adjusting their responses to electromagnetic waves after fabrication is crucial for developing more efficient and compact optical systems for a broad range of applications including sensing, imaging, telecommunications, and data storage. Chalcogenide-based phase change materials (PCMs) have shown great promise due to their stable, non-volatile phase transition between amorphous and crystalline states. Nonetheless, optimizing the switching parameters of PCM devices and maintaining their stable operation over thousands of cycles with minimal variation can be challenging. In this paper, we report on the critical role of PCM pattern as well as electrical pulse form in achieving reliable and stable switching, extending the operational lifetime of the device beyond 13,000 switching events. To achieve this, we have developed a computer-aided algorithm that monitors optical changes in the device and adjusts the applied voltage in accordance with the phase transformation process, thereby significantly enhancing the lifetime of these reconfigurable devices. Our findings reveal that patterned PCM structures show significantly higher endurance compared to blanket PCM thin films.

\end{abstract}

\section{Introduction}

\paragraph{}

Achieving post-fabrication tunability in micro-optical devices is of broad interest to applications spanning adaptive optics, beam steering, multispectral imaging, heads-up display, and beyond. Such tunability has been implemented with methods such as mechanical stretching\cite{Pryce_2010}, thermo-optic modulation\cite{miller2020large}, liquid crystals\cite{bisoyi2021liquid}, and magnetic field biasing\cite{Valente_2015} in various active materials. Among them, chalcogenide-based phase change materials (PCMs), a class of active materials exhibiting significantly different optical, electrical, and thermal properties between their crystalline and amorphous states\cite{raoux2009phase, Wuttig_2017, aryana2021interface}, have garnered significant attention due to their non-volatile phase transformation and large contrast in their refractive index ($\Delta n \sim 1-2$)\cite{aryana2023optical}. In particular, PCM-based optical metasurfaces have been shown to offer large tunability and allow for pixel level control, a feature not possible with mechanical switching\cite{shalaginov2020design}.

Metasurfaces are a class of optical components that have emerged as a driving force in modern optical technology due to their unparalleled capabilities in manipulating light in a more compact configuration. These ultrathin planar structures, composed of sub-wavelength elements, have redefined conventional optical devices by enabling precise control over light's amplitude, phase, and polarization in a remarkably compact configuration\cite{Yu_2014}. The compactness and versatility of metasurfaces have propelled them to the forefront of optical engineering, offering solutions to a myriad of challenges in various applications. Today, they are poised to revolutionize optical technology, finding applications in reconfigurable logic gates\cite{zhang2023non}, optical computing\cite{Bruckerhoff_2021}, neural networks\cite{Feldmann_2019, Shen_2017}, optical filters\cite{Sreekanth_2021, Shen_2018, julian2020reversible, popescu2023electrically}, beam steering\cite{Moitra_2022, Yin_2017, Roy_2017}, and focusing lenses\cite{Yin_2017,Roy_2017,Arbabi_2017,Wang_2016,Zhu_2015}. 

Furthermore, the ability to change the properties of the devices without a constant supply of current lowers the power requirements of PCM-based designs. These materials have a rich history, finding early application in optical discs due to their large optical contrast between amorphous and crystalline phases in response to controlled heating stimuli\cite{Wuttig_2007}. Several techniques, such as electrical\cite{Ciocchini_2016, Redaelli_2004, Ali_2020, Singh_2019} or optical pulsing\cite{Wang_2016, Michel_2014, Kolobov_2004} and thermal annealing\cite{Mandal_2021,aryana2021suppressed} have been employed to trigger phase transformation in PCMs. 

\begin{figure}[htb]
\begin{center}
\includegraphics[scale = .6]{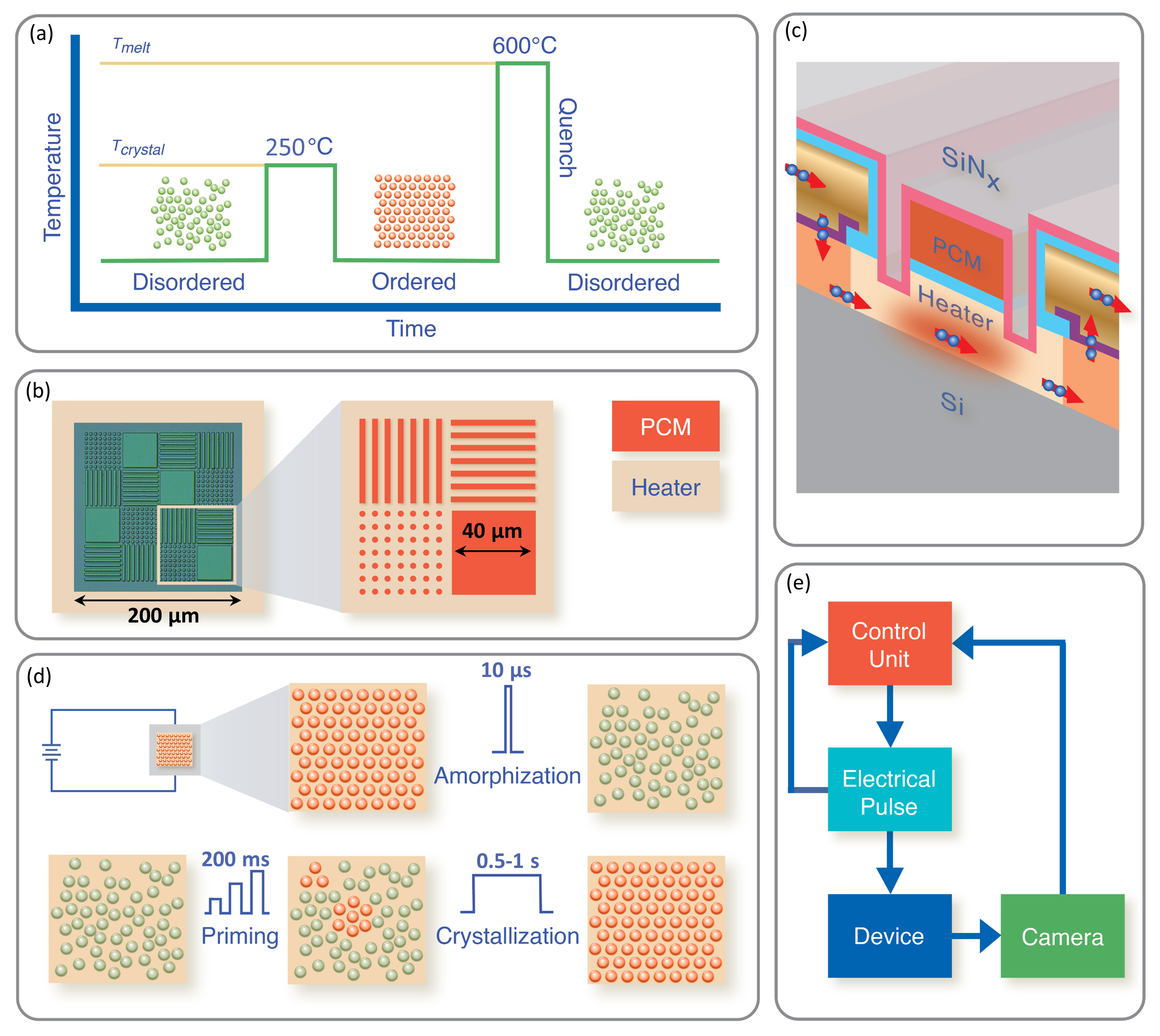}
\caption{(a) Phase transformation mechanism and corresponding pulse shape for crystallization including priming pulse to create nucleation areas \cite{orava2017classical} and amorphization pulse, (b) image of the device showing different PCM geometry on a single 200 $\upmu$m $\times$ 200 $\upmu$m microheater, (c) a cross-section schematic of the device and the corresponding layers, showing the flow of charges between the electrodes, (d) mechanism of phase transformation in PCMs and the necessary temperature rise for each switching event, (e) pulse optimization control unit.\protect\label{fig:1}}
\end{center}
\end{figure}

The integration of PCMs into micro-optical devices such as metasurfaces offers promising avenues in optical manipulation and signal processing\cite{Dong_2018, popescu2023electrically, chu2016active, yue2021nonlinear, moitra2023programmable}, holding great potential for developing compact, lightweight, and non-volatile tunable devices. This is particularly critical in space explorations, where size, weight, and power (SWaP) considerations are crucial, and highly tunable optical devices could enhance the efficiency of filtering, focusing, and steering light. Despite PCMs being predominantly utilized in small-scale devices like sub-micron memory cells, their broader application requires scaling these devices to hundreds of microns or larger. However, the process of scaling up PCM-based devices presents challenges in achieving reliable phase switching. This requires high temperatures, exceeding 600\degree C for amorphization, as shown in Fig. \ref{fig:1}(a), with fast switching timescales on the order of microseconds. As the device size increases, achieving a uniform heating profile across the PCM area becomes an intricate task\cite{Tomer_2018, aryana2023toward}. The pursuit of reliable and rapid phase switching, while mitigating thermal damage to the device, remains a complex challenge\cite{popescu2023electrically, fang2023non}.

In this study, we demonstrate robust switching of PCM-based devices through a computer-aided algorithm that controls the heating pulses with respect to the optical changes in the device, and study the impact of PCM patterns on their switching characteristics. While these devices are not metasurfaces, investigating the behavior of different geometries can provide insight into improving the performance and damage resistance of metasurfaces. For this, three device geometries were investigated: (i) 40 $\upmu$m $\times$ 40 $\upmu$m blanket-coated thin-film PCM, (ii) 40 $\upmu$m $\times$ 3 $\upmu$m and 40 $\upmu$m $\times$ 2 $\upmu$m line pattern, and (iii) 2 $\upmu$m circular dot arrays as depicted in Fig.\ref{fig:1}(b). All the geometries are deposited on IR transparent microheaters with different sizes 150 $\upmu$m $\times$ 150 $\upmu$m and 200 $\upmu$m $\times$ 200 $\upmu$m. This enables us to evaluate the performance of all the configurations under an identical operating condition. 

In order to deliver heat and trigger phase transformation, we utilize a doped-Si microheater beneath the PCM that heats up through resistive heating upon application of an electrical pulse. The cross-section of the device configuration is shown in Fig. \ref{fig:1}(c). We can control the temperature of the microheater by adjusting the pulse width and amplitude, and thus control the degree of phase transformation in PCM. In this study, we use amorphization pulses that are on the order of 10 $\upmu$s, while crystallization pulse are lower in amplitude but longer in duration, on the order of 0.5-1 s as shown in Fig. \ref{fig:1}(d). Additionally, we explore the impact of priming pulses—consisting of 6 consecutive pulses, each 200 ms wide, gradually increasing from 10 V to the crystallization voltage—on the acceleration and uniformity of crystallization within these devices \cite{orava2017classical}.  The performance of the device highly depends on the degree and quality of phase transformation, which is primarily dictated by electrical pulses delivered at the heater. As a result, it is consequential to optimize the pulses in a way that we achieve consistent phase transformation upon each cycle, while avoiding damage to the device by over-heating. Additionally, PCMs can experience drifts in their properties as they undergo thousands of switching cycles
, which causes the optical performance of the PCM-based devices to deviate from their designed states. Thus, it is important to adapt techniques to mitigate this drift in the switching property as the device cycles.

Thus far, there have been limited studies devoted to streamlining the optimization of PCM switching pulse parameters for devices beyond several tens of micrometers, which has largely relied on a trial-and-error process necessitating significant human intervention. These pulses are dependent on the geometry of the microheaters as well as its neighboring materials and thus can vary from one study to another. Manually testing the PCM-based devices to determine the optimal pulses is a time-consuming process, starting from a safe voltage and steadily increasing the amplitude or pulse width to reach maximum switching. In PCM-based devices, it is crucial to apply precisely enough energy to reach the melting temperature to avoid overheating the devices which can lead to premature failure. This process is also prone to human error during manual testing, for example causing damage upon increasing the voltage by large increments. This can drastically decrease the lifetime of the device. 

Although switching cycles in storage-class memory easily exceeds millions of cycles\cite{ding2019phase}, effectively switching PCM beyond several thousand cycles becomes increasingly challenging as the effective switching size of PCM increases, primarily due to exacerbated temperature non-uniformity across the volume of the PCM. In this paper, we strive to achieve consistent and repeatable switching events over thousands of switching cycles by focusing on the PCM pattern geometry as well as pulse optimization through a computer-aided algorithm. This technique not only automates the measurement procedure but also gives feedback to the function generator to adjust the pulses according to the device's behavior. As we will show in this paper, the automation process is critical for ensuring the reliable and reproducible operation of these devices. For this, we leverage a feedback loop that monitors the changes in the optical properties of the device using a short-wave infrared (SWIR) camera and cautiously adjusts the pulse voltage to maximize optical contrast while mitigating damage. This feedback-driven approach offers a promising solution to the challenges associated with PCM-based micro-optical devices, advancing their capabilities to new heights.

\section{Method}

\subsection{Device Fabrication}
\paragraph{}The doped silicon heaters were fabricated at the MIT Lincoln Laboratory and details of the fabrication process can be found in our prior works \cite{popescu2023open, rios2022ultra}. In short, silicon-on-insulator wafers with 1 $\upmu$m buried oxide and 155 nm of silicon were doped via ion implantation before metalization with a Ti/TiN liner and 200 nm of Al layer for electrodes. The chip was then pre-patterned via photolithography and the phase change alloy Ge$_2$Sb$_2$Se$_4$Te (GSST) \cite{zhang2017broadband, zhang2019broadband} was deposited via thermal evaporation and patterned using lift-off. The thermal evaporation was performed at a base pressure below 2 $\times$10$^{-6}$ Torr at a maximum deposition rate of $\sim$8 $\mathrm{\mathring{A}}$/s  The deposition was performed from pre-weighted source materials prepared using a melt-quench technique and crushed into powders. The target thickness was 180 nm. After the lift-off in N-methylpyrrolidone (NMP), the GSST patterns were capped at 150 \degree C in 20 nm of atomic layer deposited Al$_2$O$_3$ and further encapsulated via reactive sputtering with 800 nm of SiN$_x$. Fluorine-based reactive ion etching was used to open windows on the contact pads so the aluminum can be accessed again for wire-bonding and testing. 

\subsection{Experimental Setup}
\paragraph{}In order to monitor the device as it undergoes phase transformation, we used a SWIR camera (FLIR A6262) that operates in the wavelength range of 600 nm to 1700 nm. We used a long pass filter at 800 nm to only capture the IR portion of the signal. For magnifying the image of the microheater, we used a Navitar zoom lens with a co-axial illuminating port supported by a 10X Mitutoyo objective focusing lens. For this experiment, we monitored the changes in the reflected light from the PCM. For phase transformation, due to the large size of the device compared to PCM-based memory devices, we needed to apply voltages up to 40-50 V to trigger amorphization. For this, we used a function generator (Agilent 33220A) that was coupled with a DC power supply (Keysight E36232A) through a MOSFET switch that allowed us to deliver voltages up to 60 V. We controlled the applied voltage with a computer-aided algorithm that captured an image from the surface of the device after each switching event and analyzed the degree of the phase transformation. Additionally, several other instruments such as a multimeter and oscilloscope were used to monitor the resistance of the heater as well as the pulse shape. Ultimately, all the components and instruments worked cohesively to cycle the device and monitor its health and performance. Additional information can be found in Supporting Note 01.

\subsection{Graphic User Interface (GUI) Development and Pulse Optimization}\label{LogicInfo}
\paragraph{} The experimental procedure for studying phase transformation behavior in PCM-based devices employs an iterative process coupled with real-time data analysis to regulate pulse amplitude in relation to the extent of phase transformation. In this experiment, prior to sending each pulse, an image of the device is captured and analyzed within the software to determine the pulse amplitude in the following cycles. Starting with user-defined input parameters like pulse widths, amplitudes, and inter-pulse intervals, the software initiates the cycling process. This involves administering priming and crystallization pulses followed by an amorphization pulse to achieve a full cycle. 

The software employs a systematic analysis that relies on detecting changes in the reflectance of the PCM-covered surface on the microheater, indicative of the degree of phase transformation in PCM. Linear regression, represented by coefficients $y = m_1 x+m_2$, is used to identify trends linked to changes in reflectivity, followed by sending signals to power supply to adjust subsequent pulse amplitudes accordingly. The software updates the pulse parameters every 14 switching events, or 7 cycles. During each set of 14 switching events, the software evaluates the reflectance trends for both amorphous and crystalline states. It then adjusts the subsequent pulse based on the slope of these trends. The software adjust the pulses every 7 cycles either by only 0.25 V. This number is arbitrarily selected so that it is sufficiently small to stop damage before it propagates further into the device. The software also checks mean reflectance values and flatness of the mean switching value to identify regimes of consistent reversible switching. Once such a state is achieved, the pulse amplitudes are increased until a goal $\Delta$R is achieved. This goal for the reflectance contrast is user-specified reflectance contrast.

One crucial parameter for computer-based identification of phase transformation is the reflectance contrast, denoted as $\Delta R$, which is determined by subtracting the amplitude of the response from each pixel in the camera's image in the amorphous phase from that in the crystalline phase. The user can define a target value for how much switching (in counts) is desired per cycle. This target value generally comes from prior tests and how much change in reflectance could be achieved for a given PCM composition. If this number is too small, we would expect only partial crystallization/amorphization, while if the number is too large, the software will keep raising the voltage beyond the device tolerance and would lead to failure. Additionally, the geometry and performance requirements of the device must be considered when deciding what the target reflectance delta should be. For example, a device consisting of only the 2-D dot array pattern would achieve a lower mean switching value than one consisting of only thin films due to the lower areal fill factor of PCM.  

The algorithm assesses changes in the reflectivity of the device based on the user specified $\Delta R$ to decide whether to adjust voltage levels or maintain them. Another important parameter in the implemented algorithm is the healthy switchable area, which is defined as the area that consistently exhibits reversible changes in reflectance during cycling. The areas that are not showing any change in the reflectivity either have not been heated enough, indicating a need for more voltage to switch, or have been damaged. It is very important that the software detects damage before it propagates further and decrease the voltage accordingly. For this, we monitor the number of pixels that had already exhibited sufficient switching. Ideally, this number must increase in the beginning of the cycling process and stays flat for the reminder of the process before the damage occurs. For more details about the algorithm implementation please refer to Supporting Note 02.

\begin{figure}[htb]
\begin{center}
\includegraphics[scale = 0.55]{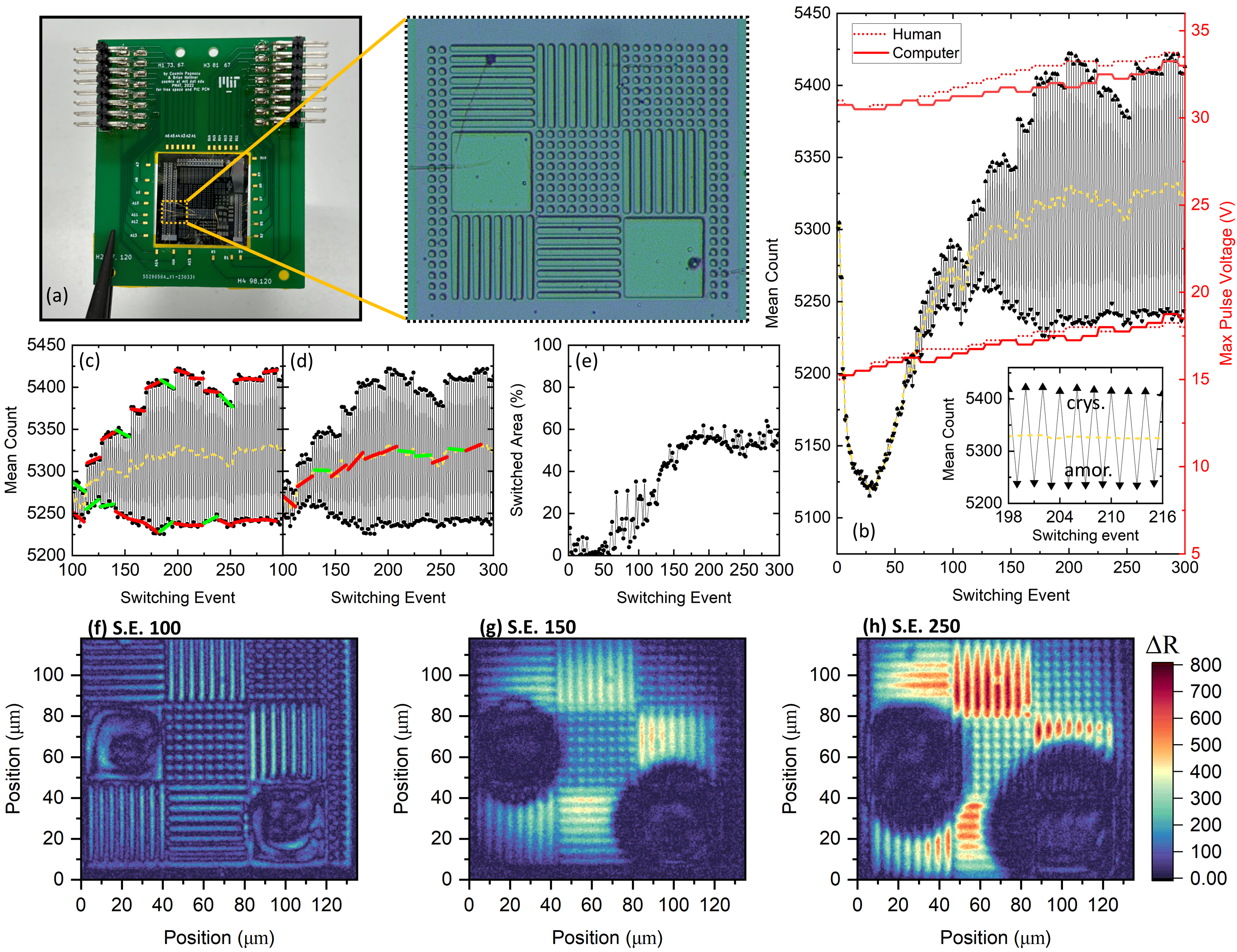}
\caption{Comparing the pulse optimization process implemented by a human operator and the adaptive algorithm on a PCM device. (a) An optical image of a chip containing PCM device arrays mounted on and wire bonded to a custom printed circuit board. The thin film, 1-D lines, and 2-D dot array PCM patterns on a 150 $\upmu$m $\times$ 150 $\upmu$m microheater are shown on the right; (b) A graph of the mean reflectance values of a PCM device, as well as the pulse voltages applied during the experiment and simulated pulses generated with the software while post-processing; (c) the conditions checked by the software to update the input parameters to ensure equal switching; (d) the condition checked by the software to update the input parameters to increase the $\Delta$R; (e) the percentage area of the device that is switching; (f), (g), and (h) the $\Delta$R between the crystalline and amorphous phases at switching events 100, 150, and 250 respectively.
\protect\label{fig:2}}
\end{center}
\end{figure}

\section{Results and Discussion}
\paragraph{}Our investigation begins with examining a microheater with dimensions of 150$ \upmu$m $\times$ 150$ \upmu$m that encompasses all three different PCM pattern geometries as depicted in Fig. \ref{fig:2}(a). Initially, we try to validate our algorithm implemented in the software by conducting tests on this device, where human operators make decisions on voltage adjustments. The results from this experiment are crucial for optimizing the control logic within the GUI and ensuring the computer can replicate human decisions accurately. To prevent damage to the device, we initiate the experiment with a conservative voltage level well below the thresholds for crystallization and amorphization. We begin the crystallization pulses at 15 V and the amorphization pulses at 31 V, gradually raising the pulse voltage until we detect reversible changes in the sample's reflectivity. To evaluate phase transformation, we capture an image from the device before each pulse and take the average reflectivity of the pixels within the region covered by PCM. This average reflectivity of the PCM-covered area as a function of switching event is depicted in Figure \ref{fig:2}(b), along with the respective applied voltages. The inset provides a closer view of the switching consistency between cycles, after the pulse voltage for crystallization and amorphization reaches sufficient threshold to trigger phase transformation. Note that the voltage is updated only after every 14 switching events. 

At the beginning of the experiment, the amorphization and crystallization pulses were not balanced, i.e. the applied pulses crystallize and amorphize PCM to a different extent, as is apparent in switching events 1-100. Specifically, the crystallization voltage proved insufficient to trigger crystallization of the PCM, while the amorphization voltage gradually increased the amorphized volume. Consequently, this leads to a decrease in reflectance below the initial uncycled state. At this point, the operator increases the crystalline voltage to trigger crystallization. The device then stabilizes in an amorphous state and the crystallization begins to catch up. Once the device stabilized again around the initial state, the operator continues to observe any trends in either pulse that reduce the reflection contrast ($\Delta$R) and increases the amplitude of that pulse accordingly. Although the experiment was conducted by a human controller, the logic outlined in section \ref{LogicInfo} was developed with the collected data. The ability of the software to detect these trends in the reflectivity in each phase is shown in Fig. \ref{fig:2}(c). The green lines indicate trends that the logic detected as lowering the $\Delta$R, while the red lines indicate that the trend was not significant enough to warrant a change in the pulse. This shows a good agreement between the decisions the human operator made and the software performance in adjusting the voltage.


After stabilizing the switching around switching event 100, the operator begins the process of increasing the $\Delta$R by increasing the amplitude of both pulses. In order to ensure that the device is switching consistently, this is only done when the mean value of each cycle is constant. Fig. \ref{fig:2}(d) shows a similar logic being implemented while post-processing the data. A linear fit was applied to the mean reflectance values of sets of 28 switching events. This longer duration was chosen to be more cautious and ensure that the device was stable. The green lines indicate regions where this criterion was satisfied and the red lines mark groups that did not meet this consistency check. These evaluations showed good agreement with the decisions the human operator had made during the experiment when trying to increase the $\Delta$R. After running these switching criteria checks in the results post-processing for this device and several other similar devices (refer to Supporting Note 03), these conditions proved to be a good representation of the input parameter changes the human operator made. The computer-generated pulses are shown in Fig. \ref{fig:2}(b) and can be compared to the human-operated pulses. The computer mimicked the human decisions at each 14th switching event closely and erred on the side of caution when it deviated.

As the device was cycled through, the switched area of the heater grew as well. The percentage area of the device switching as the experiment progressed is shown in Fig. \ref{fig:2}(e). The maximum value reached was around 50\%. Once a large $\Delta$R was achieved, the experiment was terminated after 300 switching events. A pixel was assumed to be switching if it had a $\Delta$R of at least 100 counts to avoid capturing noise in the camera as a switched region. This is a reasonable assumption considering that the color maps in Fig. \ref{fig:2}(f-h) where the areas with adequate contrast show a close representation of the actual PCM patterns. Even with the small number of switching events, the phase transformation behaviors of the three PCM patterns differ considerably. It can be seen from Fig. \ref{fig:2}(f-h) that although the change in mean reflectance for the device increases upon cycling, the unpatterned PCM thin film regions on the device begin to delaminate and the corresponding damage spreads to the other areas of the device as the device was cycled further. However, the line and dot array patterns showed much more resilience and were less prone to delamination.

The investigation into the behavior of PCM devices revealed significant differences in switching characteristics among various geometries. Notably, larger PCM regions tend to require higher switching voltages, which can be attributed to their larger thermal mass. This observation underscores the challenge of achieving uniform switching across non-uniform PCM geometries, which is particularly relevant for applications such as beam-focusing metasurfaces where precise control over switching several different PCM pattern geometries is crucial.

\begin{figure}[htb]
\begin{center}
\includegraphics[scale = 0.65]{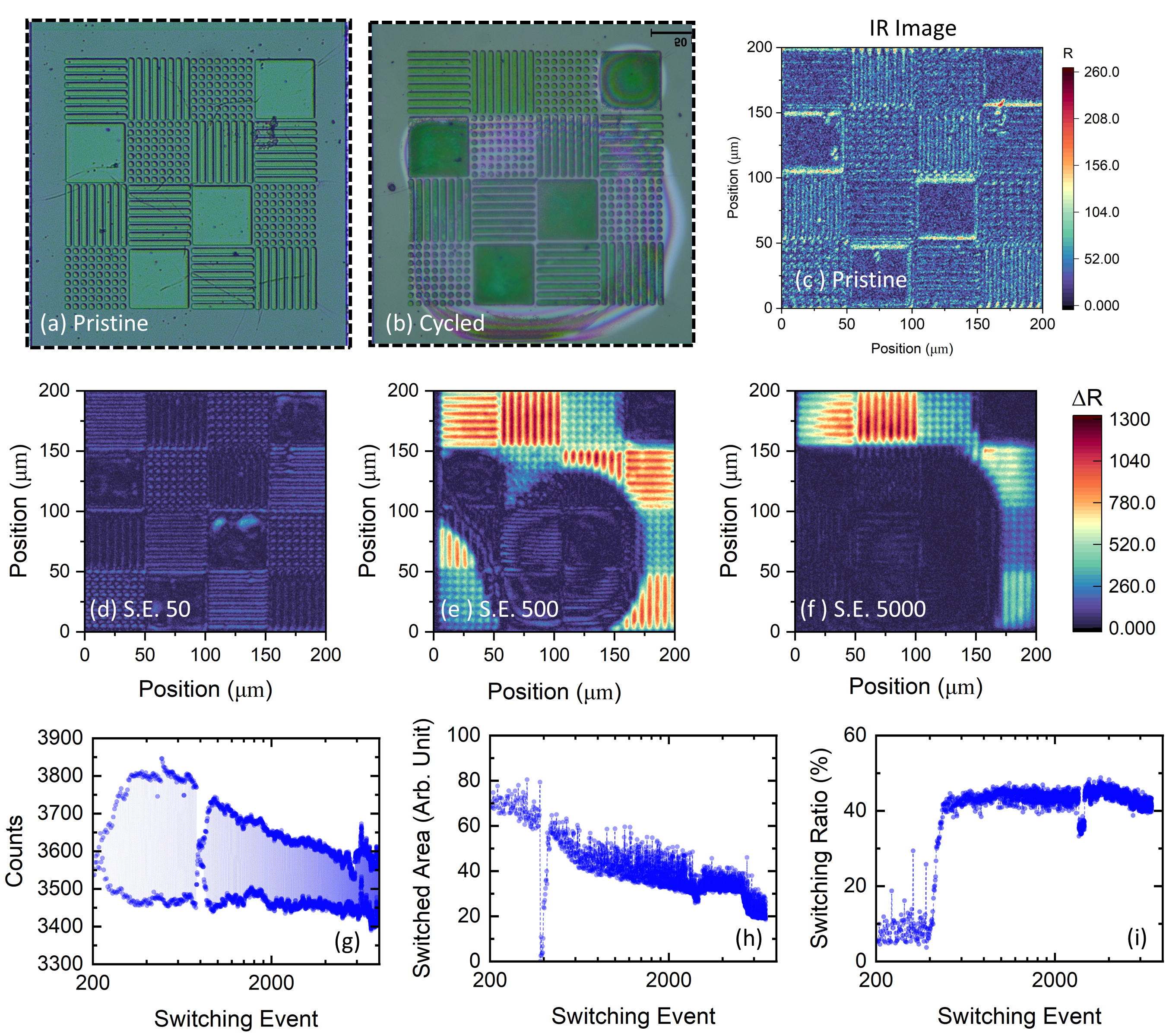}
\caption{PCM device switching performance using the adaptive pulse optimization algorithm. (a) An optical image of a device taken before the experiment featuring PCM thin film, 1-D line, and 2-D dot array geometries on a 200 $\upmu$m $\times$ 200 $\upmu$m microheater; (b) an optical image of the device taken after the experiment; (c) SWIR image of the device in its as-deposited state taken prior to the start of the experiment; (d), (e) and (f), the $\Delta$R between the crystalline and amorphous phases at switching events 50, 500, and 5000 switching events respectively; (g) the average reflectance of the device as a function of switching event; (h) the
area of the device that is switching as a function of switching event; and (i) the switching ratio of the pixel with the largest switching value.
\protect\label{fig:3}}
\end{center}
\end{figure}

Once we verified the logic for controlling the applied pulses is operating as expected, we moved to a different device to implement the adaptive pulse optimization and evaluate the performance of the device for computer-based control. For this, we use a larger microheater 200 $\upmu$m $\times$ 200 $\upmu$m, as is shown in Fig. \ref{fig:3}(a). This device consisted of 40 $\upmu$m $\times$ 40 $\upmu$m square thin film regions, 40 $\upmu$m $\times$ 2 $\upmu$m 1-D lines, and 2 $\upmu$m $\times$ 2 $\upmu$m 2-D dot arrays. The image in Fig. \ref{fig:3}(b) was taken after the cycling the device for 7000 switching events, and shows the damage on the device due to delamination. Fig. \ref{fig:3}(c) is provided as a reference for the IR image of the device prior to cycling. Figs. \ref{fig:3}(d-f) demonstrate the formation of delamination areas in a central thin film region, which propagates from the thin films to the surrounding regions. The footprint of the damage increases towards the end of the experiment respectively. 


This device was cycled for 7000 switching events, and the average device reflectivity is over the course of the experiment is shown in Fig \ref{fig:3}(g). The $\Delta$R decreased as it was cycled past 500 switching events. This trend was due to the reduction of the switching area, not a drop in the contrast in areas where delamination did not occur and the PCM continued to switch. Fig \ref{fig:3}(i) shows the percentage of the device's area that was switching, and due to the spread of the damage, the value steadily decreases. The $\Delta$R at the healthy part of the device, shown in Fig. \ref{fig:3}(i), maintains about 40\% switching for the full duration of the experiment. This indicates that if the delamination damage spread could have been mitigated, it may have been possible to sustain a large $\Delta$R for the full device.

\begin{figure}[htb]
\begin{center}
\includegraphics[width=0.95\textwidth]{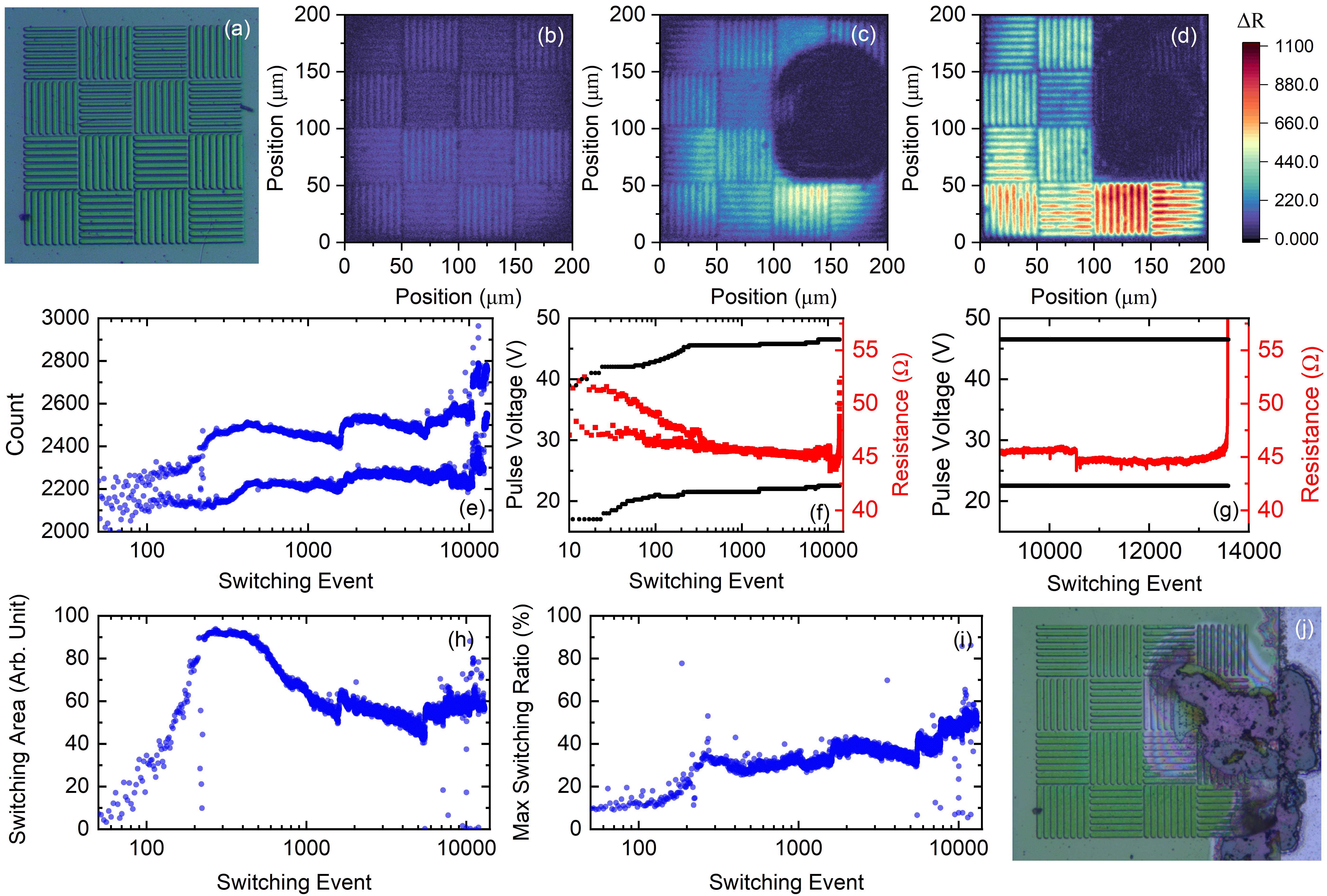}
\caption{Demonstrating enhanced endurance in a device containing only 1-D line pattern PCM geometries. (a) An image of a device featuring 2$\upmu$m (upper half) and 3$\upmu$m (lower half) gratings on a 200$\upmu$m $\times$ 200$\upmu$m microheater; (b-d) switching contrast images of the device contrast at switching events 125, 1250, and 12500 respectively; (e) the mean reflectance of the healthy PCM regions of the device as a function of switching event; (f) the prescribed voltages and measured resistances during the experiment; (g) the prescribed voltages and measured resistances at the end of the experiment (h) the device area switching as a function of switching event; (i) the ratio of the reflectivity in crystalline to amorphous phase for the pixel with the maximum switching each switching event; and (j) an image of the device taken at the conclusion of the cycling.
\protect\label{fig:4}}
\end{center}
\end{figure}

Given that patterned PCM exhibits superior endurance compared to blanket PCM films against delamination damage, we postulate that a device consisting solely of PCM patterns will significantly outperform the previously discussed devices. This is verified by our experimental results using devices that contain exclusively 1-D line patterns as illustrated in Fig. \ref{fig:4}(a), featuring dimensions of 2$\upmu$m$\times$40$\upmu$m for the upper half and 3$\upmu$m$\times$40$\upmu$m for the lower half, onto a 200$\upmu$m $\times$ 200$\upmu$m microheater. Figures \ref{fig:4}(b-d) demonstrate the contrast between images taken after crystallization and amorphization by the SWIR camera. Essentially, these images represent the difference in reflectivity of the PCM ($\Delta R = |R_{crys} - R_{amor}|$) between the crystalline and amorphous phases at switching events 125, 1250, and 12500. Brighter areas indicate a more significant change in PCM reflectivity during the switching event, which suggests a stronger phase transformation.


Similar to previous measurements, we begin the experiment with low voltage pulses and gradually raise the voltage to detect changes in the reflectivity. Figure \ref{fig:4}(e) shows the mean reflectivity as a function of switching event for the entire heater area. A video demonstrating the switching process and damage formation is available in the supplementary materials. The corresponding pulse voltages as well as the heater resistance is plotted in Figure \ref{fig:4}(f). According to our observations, the reflectivity of the PCM-covered area on the microheater remains unchanged during the first few tens of cycles. However, as the pulses reach 17 V for crystallization and 40 V for amorphization, we start to detect minor shifts in the reflectivity of the lines as shown in \ref{fig:4}(e). As we increase the voltage further, we observe a larger contrast in reflectivity between the amorphous and crystalline states. However, at switching event 399, we notice that a small portion of the sample stops switching upon phase transformation, and this region expands further during cycling, as depicted in Fig. \ref{fig:4}(c) at switching event 1250. As the device continues to be cycled, the algorithm detects slight changes in the reflectance and adjusts the parameters accordingly. These deviations were recognized by the computer after thousands of cycles of consistent switching. Detecting these would be difficult for a human, given their slow and lengthy nature, underscoring the utility of using a computer-aided algorithm to optimize and maintain the switching of the device. Towards the end of the experiment, the resistance decreased slightly but then increased drastically, as shown in Fig. \ref{fig:4} (g). This drastic increase in resistance is due to the damage to the heater. 

Figure \ref{fig:4}(h) shows the maximum switching area covered by the PCM as a function of switching event. Initially, the switching area increases gradually and peaks at approximately the 200th switching event, with over 90\% of the device area undergoing a phase transformation. Over the following 200 switching events, the switching area remains consistent before signs of damage begin to appear. As the damaged area enlarges, the maximum switching area decreases until the 1000th switching event. Following this, it maintains a relatively steady level until the device eventually fails. In Figure \ref{fig:4}(i), the maximum switching ratio ($R_c/R_a$) is depicted as a function of switching event. Here, we focus on the most substantial contrast, irrespective of the switching area, which could be as small as one pixel. This parameter signifies that in an ideal scenario where we could achieve perfect PCM switching uniformly across the entire microheater area, we would observe nearly a 50\% change in reflectance at the operating wavelength of the camera. Fig. \ref{fig:4}(j) shows an image of the device at the conclusion of the experiment. The damage in the heater due to electromigration of the metal contact is evident in the blackened regions visible on the right half of the device.

Throughout the cycling experiment, the GUI facilitated dynamic adjustments to input voltages in response to changes in the desired reflectance delta, ensuring optimal device performance. Of particular significance is the observation of PCM exhibiting good switching performance towards the end of the experiment, with a maximum optical contrast nearing 1000 counts. This shows the benefits of managing the input parameters with the GUI to avoid prematurely over-stressing the device due to human errors and impatience. This experiment was repeated for another similar pad, and the results of that test have been included in Supporting Note 04.

We can see that the patterned PCM, i.e. the lines and dot arrays, demonstrated greater resistance to damage propagation compared to unpatterned thin films. In previous assessments of PCM devices that were continuous films, damage propagation occurred rapidly following initial delamination\cite{aryana2023toward,popescu2023learning, popescu2023electrically}. In contrast, the patterned PCM demonstrated distinct damage mitigation patterns. These findings emphasize the intricate interplay between PCM geometry, switching behavior, and device durability. This improved durability may be attributed to the presence of direct contact areas between the Al$_2$O$_3$/SiN capping layer and the heater in patterned PCM regions. As a result of better adhesion between the capping layer and Si heater compared to that between PCM and heater, localized delamination in the patterned areas minimally affects neighboring regions. This highlights the potential of these geometries for applications requiring prolonged device lifetimes and reliable performance.

In conclusion, using a systematic approach, we developed a GUI to automate measurements and control pulse amplitudes, carefully pushing the device to its maximum contrast between phases while minimizing damage. Our results indicate that blanket-coated PCM films are highly prone to delamination due to weak adhesion between the PCM and heater, whereas patterned structures such as lines and dot arrays exhibit greater resilience over repeated cycles. This is attributed to the direct contact between the encapsulating layer and heater in the patterned regions, which enhances robustness and effectively prevents the propagation of damage. Notably, we achieved over 13,000 switching events for PCM line patterns on a 200 $\upmu$m $\times$ 200 $\upmu$m microheater. This underscores the benefit of using micro-structured PCM architectures such as diffraction gratings and metasurfaces for device longevity, as it provides locations where the encapsulating layer can directly bond to the heater underneath, acting as pillars to inhibit damage spread. Our findings will contribute to enhancing the reliability of electrically switchable PCM micro-optical devices toward their practical deployment.

\section{Disclaimer}
Specific vendor and manufacturer names are explicitly mentioned only to accurately describe the test hardware. The use of vendor and manufacturer names does not imply an endorsement by the U.S. Government nor does it imply that the specified equipment is the best available.

\bibliographystyle{MSP}
\bibliography{main.bbl}

\section*{Data availability} 
    The data that support the findings of this study are available from the corresponding author upon reasonable request.

\section*{Acknowledgements}
    The authors appreciate the support by Mr. Ronald Neale and Mr. Stanley H. Husch in graphic design and Mr. William Humphreys, chief engineer at NASA LaRC, for review with substantial feedback on the paper.
    Mr. Parth Garud performed his work while an intern funded through the NASA Office of STEM Engagement. Dr. Kiumars Aryana performed his work funded by the NASA Postdoctoral Program (NPP) through Oak Ridge Associated Universities (ORAU).
    This research was sponsored by the National Aeronautics and Space Administration (NASA) through a contract with ORAU and by the National Science Foundation under Awards 2132929 and 2225968. This work was carried out in part through the use of MIT.nano's facilities. The views and conclusions contained in this document are those of the authors and should not be interpreted as representing the official policies, either expressed or implied, of the National Aeronautics and Space Administration (NASA) or the U.S. Government. The U.S. Government is authorized to reproduce and distribute reprints for Government purposes notwithstanding any copyright notation herein.

\end{document}


\pagestyle{fancy}
\rhead{\includegraphics[width=3cm]{vch-logo.png}}
\setlength{\parindent}{1em}
\setlength{\parskip}{0.75em}

\title{\noindent Robust electrothermal switching of optical phase change materials through computer-aided adaptive pulse optimization}

\maketitle

\author{Parth Garud$^{1,2}$*},
\author{Kiumars Aryana$^1$*},
\author{Cosmin Constantin Popescu$^3$},
\author{Steven Vitale$^4$},
\author{Rashi Sharma$^5$},
\author{Kathleen Richardson$^5$},
\author{Tian Gu$^{3,6}$},
\author{Juejun Hu$^{3,6}$},
\author{Hyun Jung Kim$^1$}%

\begin{affiliations}
\noindent$^1${NASA Langley Research Center, Hampton, VA 23681-2199, USA}\\
\noindent$^2${Daniel Guggenheim School of Aerospace Engineering, Georgia Institute of Technology, Atlanta, GA 30332, USA}\\
\noindent$^3${Department of Materials Science and Engineering, Massachusetts Institute of Technology, Cambridge, 02139, MA, USA}\\
\noindent$^4${Lincoln Laboratory, Massachusetts Institute of Technology, Lexington, MA 02421, USA}\\
\noindent$^5${The College of Optics and Photonics, Department of Materials Science and Engineering, University of Central Florida, Orlando, FL 32816, USA}\\
\noindent$^6${Materials Research Laboratory, Massachusetts Institute of Technology, Cambridge, 02139, MA, USA}\\

\noindent
*These authors contributed equally.

\noindent
Email: kiumars.aryana@nasa.gov, hyunjung.kim@nasa.gov

\end{affiliations}

\vspace{7mm}
\begin{center}
(Supporting Information)
\end{center}
\clearpage

\def\setup{1}
\def\logic{2}
\def\tuning{3}
\def\rely{4}
\def\damage{5}

\textbf{Supporting Note \setup: Setup}

\begin{figure}[htb]
    \centering
    \includegraphics[width = 0.75\linewidth]{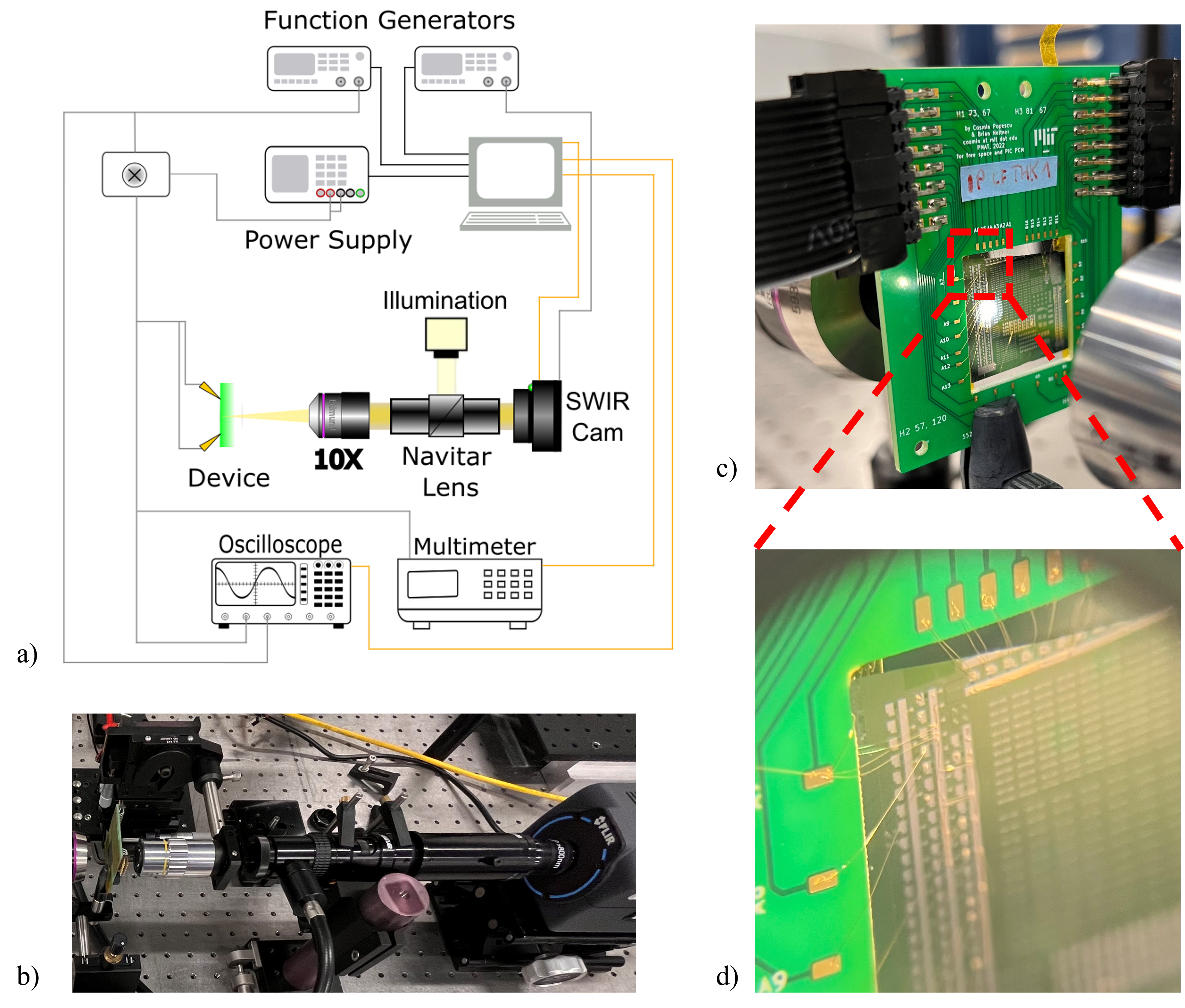}
    \caption{The experimental setup used for the testing. a) A schematic of the the full setup, with black lines representing signals and pulses and yellow lines representing feedback data; b) an image of the optical instruments set up on an optical table; c) an image of the PCB installed in the illumination path; and d) an image of the leads on the chip wire-bonded to the traces on the PCB.}
    \label{SI_setup}
\end{figure}

In order to improve the reliability of these devices, the experiment setup consisted of two distinct sections. An optical setup was required to accurately measure the transmission and reflectance of the device. This data was crucial in determining the performance of the device in-situ. An electronic setup was also required in order to accurately generate DC pulses to send to the device. Undesired fluctuations in these pulses could cause the device to not switch or prematurely fail. A schematic of the setup is shown in Fig. \ref{SI_setup}a.

The optical setup used in this experiment consisted of several carefully placed optical components. The primary illumination source in the experimental apparatus is a bulb whose light is reflected into a Navitar 5x zoom lens. This light illuminates the device being tested.

The light also is used to illuminate a short-wave infrared (SWIR) camera. Our setup utilized a A6262 FLIR camera to acquire IR images of the device at each switching event. This optical setup, shown in Fig. \ref{SI_setup}b, assures that the device is properly illuminated.

Once the device had been properly positioned in front of the objective, it was connected to the electronic instruments. The pulses were triggered by an Agilent 33220A function generator, but an external Keysight E36232A DC Power Supply was used to supply the necessary voltages since the function generator could only supply up to 5V. These two instruments worked in tandem through a trigger circuit to create and send precise high voltage pulses to the microheater under the PCM. The leads from the instruments were connected to a printed circuit board (PCB), shown in Fig. \ref{SI_setup}c. The appropriate traces on the PCB were then wire-bonded using 1 mil gold wire to the leads on the the chip for the device being tested. These connections are shown in Fig. \ref{SI_setup}d.

Additionally, several instruments monitored the device. A multimeter measured the resistance of the microheater after each pulse to ensure that it is not damaged. Furthermore, an oscilloscope was used to observe the waveforms of the pulses sent by the trigger circuit. One channel read the pulses at the function generator to ensure that pulses of the correct width are being sent at the correct time. Another channel read the voltage seen by the device. This is useful to observe the pulse shapes that are sent and can be used to identify any erroneous pulses. These instruments and connections can be seen in Fig. \ref{SI_setup}a.

\textbf{Supporting Note \logic: Pulse Optimization Logic}

\begin{figure}[h!]
    \centering
    \includegraphics[width=.85\linewidth]{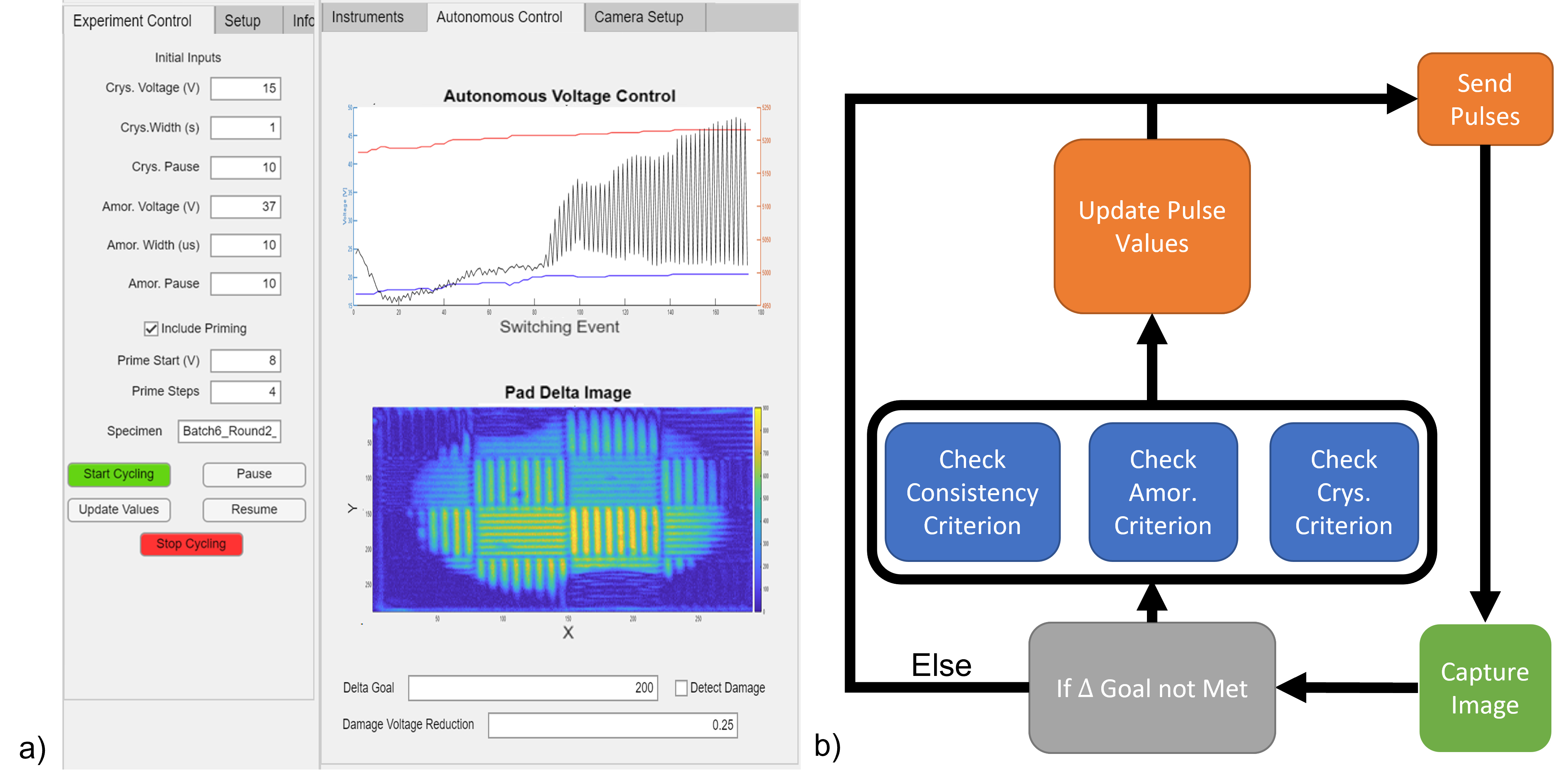}
    \caption{a) A snapshot of the experiment control and autonomous control panels of the GUI while an experiment is being run; and b) A flowchart of the high-level logic the algorithm employs.}
    \label{SI_flowchart}
\end{figure}

The graphical user interface (GUI) was developed to provide the researchers with a platform to automate the measurement process and facilitate optimization of pulse parameters. A snapshot of the GUI during the experiment is shown in Fig. \ref{SI_flowchart}a. Additionally, the software creates a log file of the nominal pulse parameters prescribed by the software or user throughout the experiment for record-keeping and test documentation. The crystallization and amorphization pulses sent to the device are optimized using feedback collected from the device in-situ. This section outlines the comprehensive experimental procedure for investigating the dynamic behavior of the opto-electronic device. The iterative approach employed in this study provides a robust framework for precise parameter control and real-time data analysis.

The experiment begins with users specifying key input parameters, including pulse widths, amplitudes, priming configurations, and inter-pulse durations. Example values for these fields can be seen in Fog. \ref{SI_flowchart}a. These values serve as the foundational settings and remain user-adjustable throughout the experiment. The experimental process involves software commands directing the DC power supply and function generators to deliver precise pulses to the device while simultaneously triggering the camera for image acquisition. Additionally, the software reads the heater's resistance via a multimeter and captures images of the device in both states with the FLIR camera.

The experiment includes an assessment of heater resistance to check the health of the microheater. A significant change in resistance suggests potential heater damage, leading to the immediate termination of the experiment. In the absence of substantial resistance changes, the heater is understood to be healthy. The software then analyses the images captured by the FLIR camera to gain insights into the device's performance. These insights are used to determine if any changes need to be made to the prescribed pulses every seven cycles. It takes several metrics into account, and a linear regression of each metric is taken to identify the overall trend and disregard noise. The software fits an equation of the form $y = m_1x + m_2$ to the data for each metric for each set of seven cycles, and the value of $m_1$ is used in the analysis. An overview of the analysis is shown in Fig. \ref{SI_flowchart}b.

The first metric checked is the damage to the PCM. Damage is detected by observing the number of pixels that demonstrate good switching as well as the number of pixels that are not switching. To achieve this, the picture of the device in the crystalline phase is subtracted from the image of the device in the amorphous phase. This subtracted image is referred to as the 'delta image'. Good switching was defined after empirical observation as showing a $\Delta$R defined by the function:

$$\Delta \mathrm{R} > \bar{c}\ \frac{hw}{A_s}$$

Where $\bar{c}$ is the average switching value of the pad, $h$ is the height of the pad in pixels, $w$ is the width of the pad in pixels, and $A_s$ is the number of pixels switching. A pixel is considered to be switching if $\Delta$R is greater than both 50 and $\hat{c}/2$, where $\hat{c}$ is the average $\Delta$R of the top 10\% best-performing pixels.

A decrease in the number of pixels exhibiting good switching could signify that the device is getting damaged. Delaminated or dewetted areas of the device would stop switching, causing the number of good switching pixels to drop. This drop condition is checked numerically by seeing if $m_1 < -10$. Furthermore, to make the regression more reliable, the two prior sets are used when finding the linear fit. This helps the software diagnose overall trends rather than local noise. 

The algorithm also determines that damage is occurring by observing the number of pixels that are not switching, defined by $\Delta \mathrm{R} < 50$. If the number of pixels not switching is increasing ($m_1 > 30$), the software diagnoses the device as being damaged by the pulses. The 50 count used in both cases was determined empirically to eliminate noise from being interpreted as switching. If either of the damage criteria are met, a signal is sent to reduce both the crystallization and amorphization pulses by a user-defined amount. This amount can be set to 0V to disable the damage detection feature. Another signal is sent to prevent any other part of the code from increasing either amplitude for that set.

The next metric that is observed is the reflectance of the device in the crystalline phase. In this phase, the device should be more reflective. Thus, an increase in reflectance is indicative of an increase in crystallization. Following this logic, if a decrease in reflectance in the crystalline state is detected, a signal is sent to increase the amplitude of the crystallization pulse by 0.25V. This signal is only sent if $m_1 < -1$ to increase the safety of this operation Furthermore, if $m_1 < -5$, the amorphization voltage is decreased by 0.25 to aid the device in trending back towards the crystalline state.

The same logic can be applied in reverse to the amorphous phase. In this phase, the device should be less reflective. Following a similar logic as the crystallization reflectance, if an increase in reflectance in the amorphous state ($m_1 > 0.5$) is detected, a signal is sent to increase the amplitude of the amorphization pulse by 0.25V. Similarly, if $m_1 > 5$, the crystallization voltage is decreased by 0.25 to aid the device in trending back toward the amorphous state.

Furthermore, the mean values of the device in both phases for each set are compared to the mean value of the first two switching events, $\bar{c}_0$. If the mean crystalline reflectance value for a set is less than $\bar{c}_0$, a signal is sent to increase the crystallization voltage by 0.25 V since the device is still mostly amorphous after the crystallization pulses have been sent. Similarly, a signal is sent to increase the amorphization voltage by 0.25 V if the mean amorphous reflectance value is greater than $\bar{c}_0$.

The final metric that is checked is the flatness of the mean switching value. This value refers to the average of the mean counts in each phase. This is done to determine if the device is trending towards either amorphous or crystalline. If this were the case, the $m_1$ coefficient would be positive (if trending crystalline) or negative (if trending amorphous). If $|m_1| < .15$, the regression is considered to be flat, and a signal is sent to increase both the crystallization and amorphization pulses by 0.25V in hopes of increasing the delta between the phases. This signal is sent independently of the signals described in the previous two paragraphs, so it is possible for each pulse to increase by up to 0.5V each set.

Essentially, if changes in reflectance in the crystalline or amorphous state are observed, suggesting a need for adjustment, the software may increase the amplitude of the corresponding pulse by 0.25 V. However, it exercises caution, only implementing adjustments if certain conditions are met to ensure device reliability. For instance, in the case of reflectance changes, adjustments are made only if the slope of linear regression on the data ($m_1$) exceeds predefined thresholds (-1 counts per cycle for crystallization and 5 counts per cycle for amorphization), minimizing the risk of over-correction. These values are derived from observing the changes made by human control and determining how steep a trend was, on average, when the human operator made a decision to increase the pulses. When this criterion was tested against other pads not used in the threshold determination process, they were increased slightly to make the computer's decisions more conservative and risk-averse. 

Additionally, mean reflectance values are compared to initial switching events, guiding further adjustments. This is done to prevent biases towards either phase. While the criteria mentioned so far are used to ensure that the device is switching consistently to the same crystalline and amorphous reflectance values to maintain an optical contrast value, additional steps are required to increase this contrast to the desired value. To accomplish this, the flatness of the mean contrast value is observed. A flat mean contrast value implies that the device is switching consistently between the two phases and the pulses sent to the device are balanced. If the slope of the linear regression of this data is below 0.15, it increases both pulse amplitudes by 0.25V each in an effort to increase the crystalline reflectance and decrease the amorphous reflectance. These incremental adjustments allow the software to fine-tune pulse parameters iteratively, aiming to optimize device performance while prioritizing reliability.

The GUI includes a switch to toggle automatic optimization on and off. If the switch is on, the GUI will execute these checks every 14 switching events to optimize the pulse parameters. It will attempt to reach a reflectance/transmission delta specified by the user and will continue to update the amplitudes of the pulses until the goal delta is reached. This goal value may depend on the application of the PCM device, the geometry of the patterns, or the size of the device.

Users have the flexibility to manually terminate the experiment using the 'Stop Cycling' function or to set an automatic termination after a predefined number of cycles using the countdown feature.

This structured and iterative experimental procedure enables precise examination of the device's dynamic behavior while maintaining flexibility and control for parameter adjustments by the user.

\textbf{Supporting Note \tuning: Algorithm Tuning}

\begin{figure}[h!]
    \centering
    \includegraphics[width=.8\linewidth]{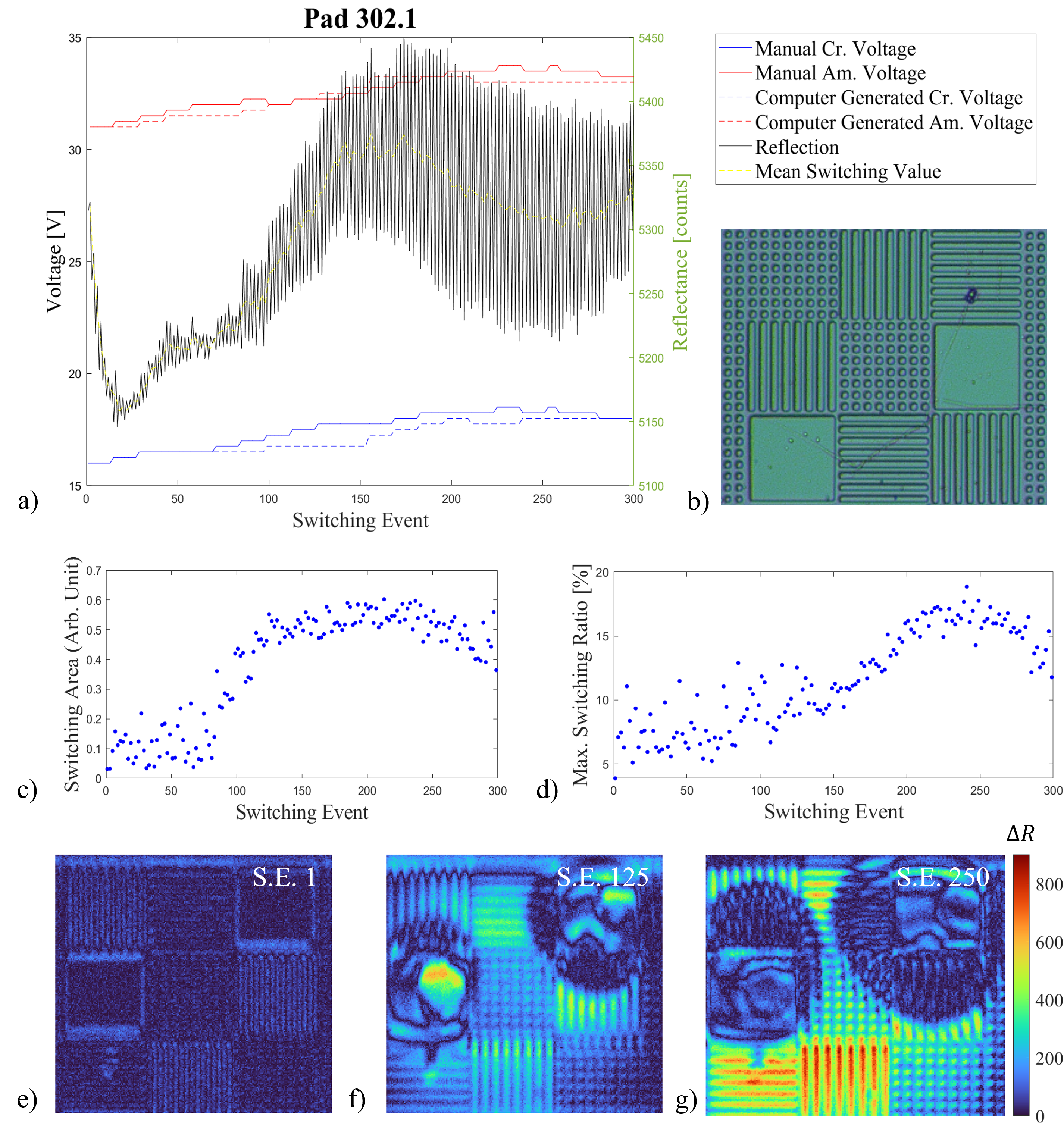}
    \caption{A device used for algorithm tuning. a) the measured reflectance, prescribed inputs, and computer generated pulses; b) an optical image of the device in its uncycled state; c) the switching area of the device as a function of switching event; d) the maximum switching ratio, $R_C / R_A$, as a function of switching event; and e), f), and g), the device contrast for switching events 1, 125, and 250 respectively.}
    \label{SI_1}
\end{figure}

Several 150 $\upmu$m $\times$ 150 $\upmu$m pads were used to tune the thresholds specified in Supplementary Note \logic. These devices were manually cycled with systematic changes to the pulse parameters every 14 switching events. The trends seen in the data were correlated to the human decision that was made in order to identify criteria that could repeatably implemented on other devices. In addition to the device discussed in the main manuscript, the data gathered from a different device is shown in Fig \ref{SI_1}.

Fig. \ref{SI_1}a shows the measured reflectance of the device as its pulse parameters were optimized. The device contained 40 $\upmu$m $\times$ 40 $\upmu$m thin films, 2 $\upmu$m $\times$ 40 $\upmu$m line patterns, and 2 $\upmu$m $\times$ 2 $\upmu$m dot patterns. Initial voltages of 16V and 31V were chosen for crystallization and amorphization pulse amplitudes respectively. This starting amorphization pulse was much more effective than the starting crystallization pulse. This can be observed in switching events 1-14 with the average reflectance decreasing steeply. The crystallization pulse amplitude was then increased and and the device reflectance in the crystalline phase began to trend upward. This pattern was continued until the device reached a state where increasing the crystallization voltage did not increase the reflectance. Throughout this process, the thin film regions delaminated, but this damage was deemed acceptable in an effort to switch the more resilient line and dot patterns. Once the pulses were switching the device at a consistent $\Delta$R, the pulses were increased after switching 168 to attempt reaching a larger $\Delta$R. This was done until switching event 210, at which point any increased voltages yielded diminishing returns due to them damaging the device. The device before cycling can be seen in Fig. \ref{SI_1}b.

Fig \ref{SI_1}c highlights the initial pulse balancing phase of the experiment, where the goal was to switch the full device. If one pulse was having a larger effect than the other, the device would exhibit lower a lower switching area due to the device 'resting' in one phase. As the pulses balance out, more of the device start to switch, which can be seen around switching event 100. This initial process is referred to as the 'pulse balancing' process. At this point, the objective transitioned to increasing the average $\Delta$R. To visualize this, the maximum switching ratio is shown in Fig. \ref{SI_1}d. When both pulses were increased to attempt this, the maximum switching ratio increased until around switching event 250. This process is referred to as the $\Delta$R-optimizing process. Any increases in voltage increased the maximum switching ratio, but decreased the switching area of the device, indicating that the additional $\Delta$R at the healthiest part of the device was occurring at the expense of spreading the existing damage. 

These damage propagation patterns can be observed in Figs. \ref{SI_1}e-g. The device starts out with no switching, apparent in Fig. \ref{SI_1}e. By the time the pulses are balanced at switching event 125, the thin film regions have delaminated but the other patterns on the device are showing signs of switching, as displayed in Fig. \ref{SI_1}f. As the $\Delta$R is increased, the damage spreads due to the higher prescribed voltages. However, the undamaged regions of the device show a contrast nearly double that of what was seen when only balancing pulses.

The behavior of this device is consistent with the results for the 150 $\upmu$m $\times$ 150 $\upmu$m device presented in the main manuscript. The data from these devices and 2 other similar devices were analyzed to determine how strong a trend had to be in order to prompt a human operator to change the pulse parameters. Linear regressions were applied to each measured metric for each set of 14 switching events. The slopes of these regressions were then cross-referenced with the decisions the operator made during the experiment, and action threshold values were determined. As a safety measure, the threshold values were chosen to be slightly more extreme than the slopes that the human acted on. This was done for general conservatism and to make the algorithm more risk-averse than a human operator. The final values that were implemented into the software are included in Supporting Note \logic.

\textbf{Supporting Note \rely: Reliability Testing Results}

\begin{figure}[h!]
    \centering
    \includegraphics[width=.8\linewidth]{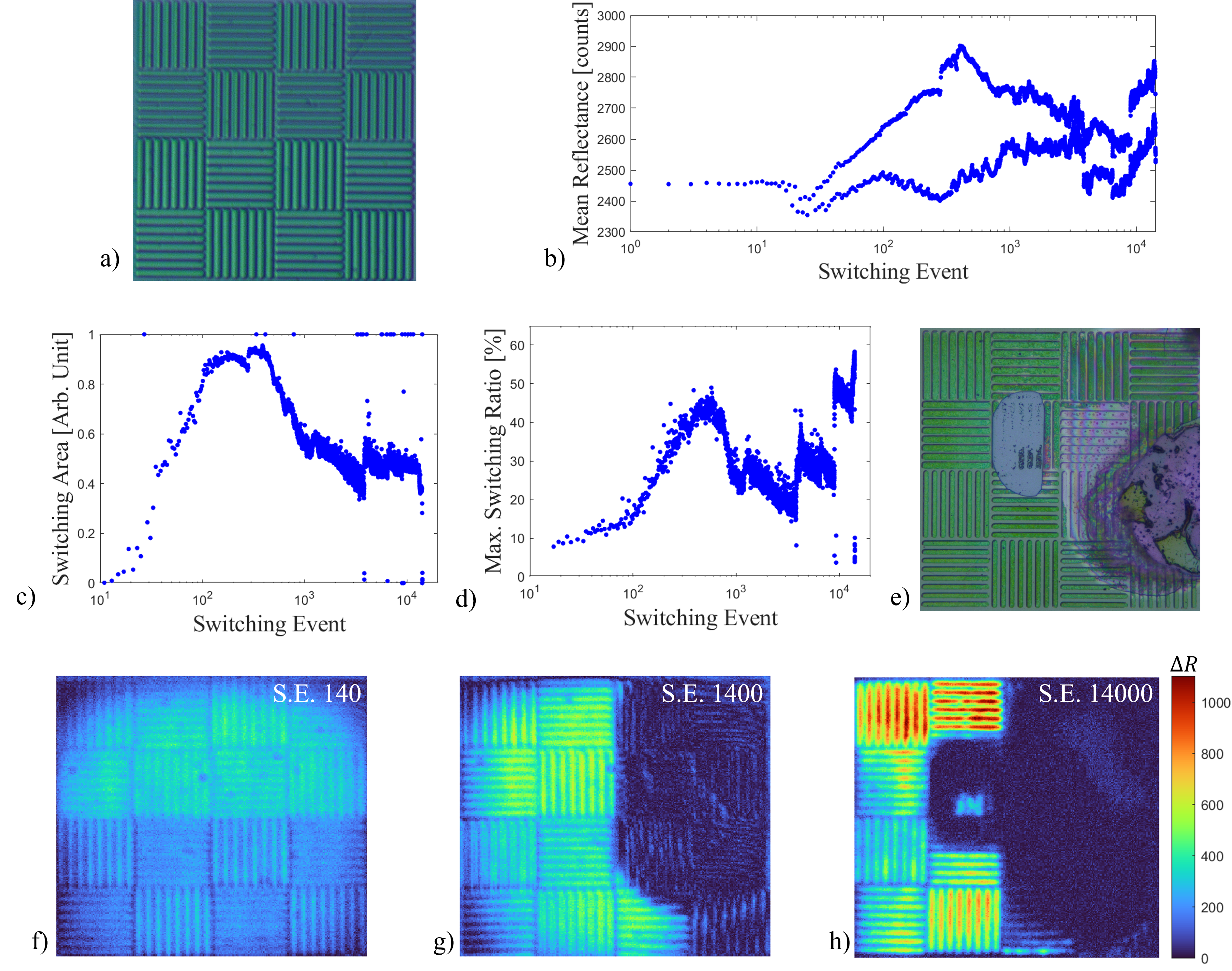}
    \caption{A device that was cycled for over $10^4$ switching events. a) an optical image of the device with 2 $\upmu$m $\times$ 40 $\upmu$m and 3 $\upmu$m $\times$ 40 $\upmu$m line patterns in its uncycled state; b) the switching area of the device as a function of switching event; c) the maximum switching ratio, $R_C / R_A$, as a function of switching event; d) the measured mean relfectance of the full device, as well as the mean reflectance of the undamaged areas of the device; e) an optical image of the device after the experiment; and f), g), and h), the device contrast for switching events 140, 1400, and 14000 respectively.}
    \label{SI_2}
\end{figure}

By implementing the logic outlined in Supporting Note \logic $\ $into a Graphical User Interface (GUI) and tuning the parameters with the testing outlines in Note \tuning, the GUI was used to cycle a device with 2 $\upmu$m $\times$ 40 $\upmu$m and 3 $\upmu$m $\times$ 40 $\upmu$m line patterns. The GUI was able to optimize the pulse parameters and monitor the performance of the device, and adjusted the optimized parameters to reflect and changes seen in the device's behavior. To supplement the device showcased in the main manuscript, the data from an additional device that was also cycled for over $10^4$ switching events is shown in Fig. \ref{SI_2}.

The microscopy image of the device is shown in Fig. \ref{SI_2}a. The 2 $\upmu$m $\times$ 40 $\upmu$m line patterns are on the upper half of the device, and the 3 $\upmu$m $\times$ 40 $\upmu$m line patterns are on the bottom half of the device. Fig \ref{SI_2}b shows the measured average reflectance of the device. This device also consisted of the pulse balancing and $\Delta$R maximizing processes outlined in Supporting Note \tuning. These phases lasted a total of around 300 cycles, consistent with the devices used to tune the algorithm. After this point, the software monitored the performance of the device and made minor changes in the pulse parameters as needed. The jumps in the data towards the end of the experiment can be attributed to stopping and restarting the software.

The trend in the switching area of the device is shown in Fig. \ref{SI_2}c. The switching area steadily increases as the pulse amplitudes are increased from the safe starting values to the balanced pulse values. This is followed by the switching ratio trend seen in Fig. \ref{SI_2}d. The switching ratio starts increasing after switching event 100, once the pulses are mostly balanced. This order of trends seen - an increase in switching area followed by an increase in the maximum switching ratio - was seen in the other devices as well. The decrease in switching area in order to achieve a higher switching ratio was also observed in the other devices. Although preliminary, this would imply that a device can be optimized for either parameter. In order to increase the switching ratio, the pulse voltages must be increased. However, the device is more likely to get damaged at these higher voltage.

The device underwent noticeable damage due to a combination of delamination and electromigration (see Supporting Note \damage). Fig. \ref{SI_2}e is an image of the device after the experiment taken with an optical microscope. Evidence of both types of damage is visible on the right half of the device. The growth of this damage can be seen in Figs. \ref{SI_2}f-h. Fig \ref{SI_2}e shows the contrast in the device partway through pulse optimization process. In this phase, the upper half with thinner line patterns has begun to switch but the thicker lines on the lower half are still not switching fully. After the optimization phase, the device enters the maintenance phase. The primary objective in this phase is to ensure that the $\Delta$R of the device does not decrease and damage spread is mitigated where possible. In Fig. \ref{SI_2}g, it can be seen that the right half of the device has delaminated, but the left half of the device is switching well. The spread of this delamination was mitigated by the software. This reduced spread, along with the effects of electromigration in the device can be seen in Fig. \ref{SI_2}h.
